\documentclass[12pt]{article}

\usepackage{amsmath, amsthm, amscd, amsfonts, amssymb, longtable, graphicx}
\usepackage[utf8]{inputenc}

\usepackage{xcolor}
\usepackage[margin=1in]{geometry} %one inch margins
\usepackage{setspace}
\title{An alternative approach to the exact network reliability assessment through the quickest path}
\date{}

\begin{document}
\maketitle

\begin{center}
\textbf{Majid~Forghani-elahabad\footnote{Corresponding author: Email address:~m.forghani@ufabc.edu.br, Phone: +55(11)49968330}}\\
Center of Mathematics, Computing, and Cognition - Federal University of ABC, Santo André, SP, Brazil	
\end{center}

		\begin{abstract}
			Extending the quickest path problem to the network reliability, a new problem emerged which aims to assess the network reliability for transmitting at least $d$ units of data from a source node to a sink node through one minimal path (MP) within a given $T$ units of time. Many of the proposed approaches in the literature check all the MPs of the network for doing the job and then construct desired system state vectors based on the accepted MPs. Hence, they need to have all the MPs of the network in advance. Here, we propose a simple approach that does not need any MP in advance. The algorithm is shown to be corrected and is illustrated through an example. \\			
			\textit{Keywords:} Network flow reliability, Quickest path reliability,  Algorithms.
		\end{abstract}

	\section{Introduction}
	A multi-state flow network (MFN) is a network in which the arcs (and possibly nodes) may have more than two states~\cite{forghani2014ieee, mansourzadeh2014comparative, forghani2015ress, yeh2021novel, forghani2015ieee, yeh2022new}. The different states of each arc show the possible capacities of the arc, which are stochastic. When the network is dynamic, each arc in the network has another attribute, the so-called \textit{lead time}, that refers to the needed time for transmitting one unit of data from the source node to the sink node~\cite{lin2003extend, forghani2022quickest, yeh2015fast, forghani20223, niu2021reliability, forghani2013simple}. The quickest path problem (QPP), a variant of the shortest path problem, has been an attractive problem in the literature~\cite{chen1990quickest}. Extending the QPP to the network reliability, a new problem called the quickest path reliability problem emerged in~\cite{lin2003extend}. 
	The problem aims at reliability evaluation of the MFN for transmitting at least $d$ units of data through one minimal path (MP) from the source to the sink nodes within a given $T$ units of time~\cite{forghani20223, yeh2009simpleQPR}. Several researchers have studied this problem and proposed different algorithms to address it so far~\cite{forghani2015ress, forghani2015ieee, lin2003extend, forghani2022quickest, forghani2013simple, yeh2015fast, forghani2022quickest, forghani20223, yeh2009simpleQPR, el2016efficient, forghani2020ijor}. However, the proposed approaches in the literature usually need all the MPs as input. Here, we propose a simple approach that does need any MP in advance. The rest of the paper is organized as follows. Some preliminaries and the proposed approach are provided in Section~\ref{mainblock}. An illustrative example is given in Section~\ref{example}, and finally, we conclude the work in Section~\ref{conclusions}.
	
	\section{Main block}\label{mainblock}
	Let $ G(N, A, M, L) $ be a multi-state flow network (MFN), where  $ N= \{1, 2, \cdots , n\} $ is the nodes' set,  $ A = \{a_1, a_2, \cdots, a_m\} $ is the arcs' set,  $ M = (M_1, M_2, \cdots, M_m) $ is a vector with $M_i$  denoting the maximum capacity of arc $a_i$, for $i=1, 2, \cdots, m$, and  $L=(l_1,l_2,\cdots,l_m)$ is a vector with $l_i$ denoting the lead time of arc $a_i$, for $i=1, 2, \cdots, m$. Hence, the numbers $n$ and $m$ are respectively the number nodes and arcs in the network. Let node $ 1 $ be the source and node $ n $ be the sink node in the network. 
	Let $x_i$ denote the current capacity of arc $a_i$, with the values from $\{0,1,\cdots,M_i\}$, for $i=1, 2, \cdots, m$, and hence $X=(x_1, \cdots, x_m)$ be the current system state vector (SSV) of the MFN.
	To each node $a_i$, for $i=1, 2, \cdots, m$, we assign two numbers;  $I_i$, which is the number of incoming arcs to it, and $O_i$, which is the number of outgoing arcs from it. As nodes $1$ and $n$ are respectively the source and the sink nodes, we have  $I_1=O_n = 0$, and as the considered MFNs are undirected, we have $I_i=O_i$, for $i=2, 3, \cdots, n-1$. 
	A path is a set of adjacent arcs which connects the nodes $1$ and $n$, and a minimal path (MP) is a path whose any proper subsets is a path. 
	Let $h$ be the number of all the MPs in the network. For minimal path of $P_j$, let $LP_j$ be its lead time and $CP_j(X)$ be its capacity under $X$. Now, letting $ P_j=\{a_{j_1}, a_{j_2}, \cdots, a_{j_{nj}}\} $, the lead time is calculated using the following equation~\cite{forghani20223}.
	\begin{align}\label{lpj}
		LP_j = \sum_{r=1}^{nj} l_{j_r}.
	\end{align}
	And letting $ X = (x_1, x_2, \cdots, x_m) $, the capacity is calculated using the following equation~\cite{forghani2019ress}.
	\begin{align}\label{cpj}
		CP_j(X)~=~\min\{x_{j_1}, x_{j_2}, \cdots, x_{j_{nj}}\}. 
	\end{align}
	
	The network reliability in this problem, denoted by $R_{d,T}$,is calculating the probability of transmitting $d$ units of data from node $1$ to node $n$ through one MP within the given $T$ units of time. Let $\rho(X,d)$ be the required time to send $d$ units of data through the quickest MP from node $1$ to node $n$ in the network, $ \Theta(d, T) = \{ X\leq M\ |\ \rho(X,d)\leq T\} $, and $ \Theta(d, T)_{\min} = \{X^1, X^2, \cdots, X^{\sigma}\} $ be the set of all the minimal solutions in $\Theta(d, T)$. Let also $A_r=\{X|X\geq X^r\}$, for $r=1,2,\cdots,\sigma$, $B_1=A_1$, and $B_r = A_r-\cup_{j=1}^r A_j$, for $r=2, 3, \cdots, \sigma$. Then, the network reliability can be calculated using the following equation~\cite{forghani2016amm}.
	\begin{align}\label{reliab}
		R_{d,T}=\Pr\cup_{r=1}^\sigma A_r=\Pr\cup_{r=1}^\sigma B_r=\sum^\sigma_{r=1}\Pr(B_r), 
	\end{align}
	where $\Pr(B_r)=\sum_{X\in B_r} \Pr(X)$ and $\Pr(X)= \prod^m_{i=1}\Pr(x_i)$. 
	
	Therefore, one needs to determine the sets $B_r$, for $r=1, 2, \cdots, \sigma$ for calculating $R_{d,T}$, for which the solutions $X^1, X^2, \cdots, \sigma$ are required. Hence, this work focuses on the determination of all such solutions. To calculate $\ \rho(X,d)$ for any SSV $X$, one needs to first calculates the transmission time of sending $d$ units of data through an arc and then through a path. 
	To send $ d $ units of data through $ a_i $ with capacity of $ x_i $ and the lead time of $ l_i $, the transmission time is equal to~\cite{forghani2022quickest}
	\begin{align}\label{arc-lead-time}
		t= l_i+\lceil \frac{d}{x_i}\rceil,
	\end{align}
	where $ \lceil \alpha \rceil $ is the first integer number greater than or equal to $ \alpha $. And to send $ d $ units of data through  $P_j= \{a_{j_1}, a_{j_2}, \cdots, a_{j_{nj}}\}$ under $X$, the transmission time is equal to~\cite{forghani20223}
	\begin{equation}\label{path-lead-time-formula}
		\varrho(P_j, X,d)=LP_j+\lceil \frac{d}{CP_j(X)}\rceil.
	\end{equation}
	According to the above equation, if $T>LP_j$, the minimum capacity of MP, $P_j$, for being possible to transmit $d$ units of data through it within $T$ units of time is equal to
	\begin{align}\label{correspondingcapcity}
		\eta_j = \lceil \frac{d}{T -LP_j}\rceil
	\end{align}
	Therefore, if $CP_j\leq \eta_j$ and one sets the capacity of all the arcs in $P_j$ to $\eta_j$ and the capacity of the other arcs to zero, then a minimal SSV is at hand, which belongs to $\Theta(d, T)_{\min}$. We note that as the data can be sent through one MP from the source node to the sink node, we have $\rho(X,d) = \min_{j=1:h}\varrho(P_j, X,d)$.
	
	Therefore, one needs to check these conditions for each MP, determine the acceptable MPs, and calculate the corresponding SSVs. However, this approach requires all the MPs as input which is a weakness. To prevent this requirement, here we use the node-child matrix of the network introduced in~\cite{fathabadi2009determining}. Our proposed algorithm directly determines only the MPs that satisfy the required conditions without duplicates and then calculates the corresponding SSVs, which are the final solutions.
	Letting $n$ be the number of nodes in the network, the node-child matrix is an $n\times q$ matrix, where $q=\max\{O_i\ |\ i = 1, 2, \cdots, n-1\}$. Each row in the matrix corresponds to a node in the network and shows the children of that node. To have a better understanding, the node-child matrix of Fig.~\ref{fig1:benchmark} is given below.
	\begin{equation}\label{matrixB}
		\textbf{B}=\begin{bmatrix}
			2 & 3 & 4\\
			3 & 4 & 0\\
			2 & 4 & 0\\
			0 & 0 & 0\\
		\end{bmatrix}
	\end{equation}
	\begin{figure}
		\centering
		\includegraphics[width=0.3\linewidth]{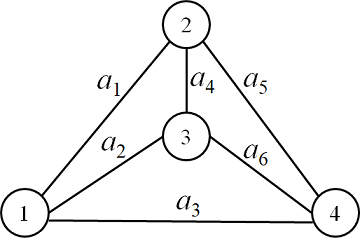}
		\caption{A benchmark network example taken from~\cite{forghani2019ijrqse}.} 
		\label{fig1:benchmark}
	\end{figure}

	With the node-child matrix of an MFN, one can determine all the MPs of the network using a backtracking procedure~\cite{fathabadi2009determining}. The proposed algorithm here adds one more condition to this procedure for checking the lead time of the under-construction MP to find all the solution vectors in the set of $\Theta(d, T)_{\min}$. If the lead time of the under-construction MP passes $T$, the algorithm stops the construction and goes back to construct the next MP. 
	One notes that any obtained solution should be less than or equal to $M$. Therefore, the algorithm checks this condition as well. Note that $L(s,t)$ is the lead time of the arc between nodes $s$ and $t$.\\
	
	\textbf{The proposed algorithm}\\
	\textbf{Input}: $ G(N, A, M, L) $ with demand level $d$ and time limit $T$.\\
	\textbf{Output}: The set $\Theta(d, T)_{\min}$.\\
	
	\textbf{Step 0.} Let $\Theta(d, T)_{\min} = \{\},\ i= 1, s=1, k=1,\ f(r) = 1 $ for $ r = 1, \cdots, n $, $lt = 0$, $kap = \infty$, $Rp=0$, $P = (1, 0, \cdots, 0)_{1\times n} $ and $ K = (0, 0, \cdots, 0)_{1\times n} $.
	
	\textbf{Step 1.} Determine the node-child matrix $ B $.
	
	\textbf{Step 2.} Let $ t=B(s, f(s)) $.
	
	\textbf{Step 3.} If $ t\in P$, then let $Rp =1$.
	
	\textbf{Step 4.} If $ t \neq 0 $, then go to Step~7.
	
	\textbf{Step 5} If $ s=1 $, then stop. $ \Theta(d, T)_{\min}$ is the set of all the solutions. Else if $  s= n $, then let $ X^k = (0, 0, \cdots, 0)_{1\times m} $, and update $x_i=\eta$,if $a_j\in P$. Let $\Theta(d, T)_{\min} = \Theta(d, T)_{\min}\cup \{X^k\}$. Now, if $ i=2 $, then  stop. Otherwise, let $ k=k+1,\ f(P(i-1))=1 $, $lt = lt - L(P(i-1), P(i))-L(P(i-2), P(i-1))$, 
	$P(i) =0,\ P(i-1) =0,\ K(i-1)=0, K(i-2)=0,\ s = P(i-2)$, and $i=i-2$. Now, if $ i=1 $, let $ kap = \infty $, else $ kap = \min\{K(j)\ |\ j=1, \cdots, i-1\} $. Go to Step~2.
	
	\textbf{Step 6} Let $ f(s) =1,\ lt = lt - L(P(i-1), P(i)),\ P(i) =0,\ K(i-1)=0,\ s = P(i-1)$, and $i=i-1$. Now, if $ i=1 $, let $ kap = \infty $, else $ kap = \min\{K(j)\ |\ j=1, \cdots, i-1\} $. Go to Step~2.
	
	\textbf{Step 7.} If $Rp=1$, then let $ f(s) = f(s)+1 $, $Rp=0$, and go to Step~2.
	
	\textbf{Step 8.} If $ lt+L(s,t)< T $, then let $ \eta = \lceil \frac{d}{T-lt-L(s,t)}\rceil $. If $ lt \geq T-L(s,t) $ or $ \eta>\min\{kap, M(s,t)\} $, then let $ f(s) =f(s)+1 $, else let $ lt = lt +L(s,t), \ kap  = \min\{kap, M(s,t)\}, \ K(i) = M(s,t), \ f(s) = f(s)+1,\ i = i+1,\ P(i) = t$, and $ s = t $. Go to Step~2.\\
	
	The above algorithm checks all the required conditions for the under-construction MPs to assure that only the MPs, say $P_j$, are constructed that satisfy $LP_j<T$ and $\tau(P_j, d, M)\leq T$. Moreover, the algorithm checks each calculated solution for being less than or equal to $M$. As a result, as the proposed algorithm here is based on the proposed algorithm in~~\cite{fathabadi2009determining}, its correctness is demonstrated, and it determines the set $\Theta(d, T)_{\min}$ correctly. We solve a benchmark example in the next section to illustrate the proposed algorithm.
	
	\section{An illustrative example}\label{example}
	Consider the given network in Fig.~\ref{fig1:benchmark} with $M=(5, 4, 6, 4, 3, 6)$ and $L=(4, 4, 1, 4, 3, 1)$ and determines the set $\Theta(4, 7)_{\min}$ by using the proposed algorithm here. 
	
	\noindent{\textbf{Solution:}}
	
	{Step 0.} Let $\Theta(4, 7)_{\min} = \{\},\ i= 1, s=1, k=1,\ f(r) = 1 $ for $ r = 1, 2, 3, 4 $, $lt = 0$, $kap = \infty$, $Rp=0$, $P = (1, 0, 0, 0) $ and $ K = (0, 0, 0, 0) $.
	
	Step~1. The matrix $B$ is calculated as given in Eq.~\eqref{matrixB}.
	
	Step~2. Let $t = B(s, f(s)) = B(1, 1) = 2$.
	
	Step~3. $t=2\notin P$.
	
	Step~4. $t=2\neq 0$. The transfer is made to Step~7.
	
	Step~7. $Rp=0\neq 1$.
	
	Step~8. Since $lt+L(1,2) = 0+4 = 4<7$, let $ \eta = \lceil \frac{4}{7-0-4}\rceil = 2 $. 
	Since $ lt =0 < 3 = T-L(1,2) $ and $ \eta = 2 < 5=\min\{kap, M(1,2)\} $, let $ lt = lt +L(1,2) = 0+4=4$, $kap  = \min\{kap, M(1,2)\}=5$, $K(1) = M(1,2)=5$, $f(1) = f(1)+1=2$, $i = i+1=2$, $P(2) = 2$, and $ s = 2 $.
	The transfer is made to Step~2.
	
	Step~2. Let $t = B(2, f(2)) = B(2, 1) = 3$.
	
	Step~3. $t=3\notin P$.
	
	Step~4. $t=3\neq 0$.
	
	Step~7. $Rp=0\neq 1$.
	
	Step~8. $lt+L(2,3) = 4+4 = 8 \nless 7$. 
	Since $ lt =4 \geq 3 = T-L(2,3) $, let $ f(2) = f(2)+1 = 2 $.
	The transfer is made to Step~2.
	
	\vspace{5mm}
	\vdots
	\vspace{5mm}
	
	The final solutions are $\Theta(4, 7)_{\min} = \{$(0, 0, 1, 0, 0, 0), (0, 2, 0, 0, 0, 2)$\}$.\\

	\section{Conclusions}\label{conclusions}
	We proposed a simple efficient algorithm to address multi-state flow networks' quickest path reliability problem. We showed the correctness of the algorithm and illustrated it through a benchmark network example. The main advantage of our proposed algorithm to the several existing approaches is that our algorithm does not need any MP as input.
	
	\section*{Acknowledgement}
	
	The author thanks CNPq (grant 306940/2020-5) for supporting this work.
	
	%\bibliographystyle{elsarticle-num}
	%\bibliography{C:/Users/Majid/Dropbox/majidbibliography}
	%\bibliography{majidbibliography}

\begin{thebibliography}{00}
		\bibitem{forghani2014ieee}
		M. Forghani-elahabad, N. Mahdavi-Amiri, A new efficient approach to search for all multi-state
		minimal cuts, IEEE Transactions on Reliability 63 (1) (2014) 154–166.
		
		\bibitem{mansourzadeh2014comparative}
		S. M. Mansourzadeh, S. H. Nasseri, M. Forghani-elahabad, A. Ebrahimnejad, A comparative
		study of different approaches for finding the upper boundary points in stochastic-flow networks,
		International Journal of Enterprise Information Systems (IJEIS) 10 (3) (2014) 13–23.
		
		\bibitem{forghani2015ress}
		M. Forghani-elahabad, N. Mahdavi-Amiri, An efficient algorithm for the multi-state two sep-
		arate minimal paths reliability problem with budget constraint, Reliability Engineering \&
		System Safety 142 (2015) 472–481.
		
		\bibitem{yeh2021novel}
		W.-C. Yeh, Z. Hao, M. Forghani-elahabad, G.-G. Wang, Y.-L. Lin, Novel binary-addition
		tree algorithm for reliability evaluation of acyclic multistate information networks, Reliability
		Engineering \& System Safety 210 (2021) 107427.
		
		\bibitem{forghani2015ieee}
		M. Forghani-elahabad, N. Mahdavi-Amiri, A new algorithm for generating all minimal vectors
		for the q smps reliability problem with time and budget constraints, IEEE Transactions on
		Reliability 65 (2) (2015) 828–842.
		
		\bibitem{yeh2022new}
		W.-C. Yeh, S.-Y. Tan, M. Forghani-elahabad, M. El Khadiri, Y. Jiang, C.-S. Lin, New binary-
		addition tree algorithm for the all-multiterminal binary-state network reliability problem, Re-
		liability Engineering \& System Safety 224 (2022) 108557.
		
		\bibitem{lin2003extend}
		Y.-K. Lin, Extend the quickest path problem to the system reliability evaluation for a
		stochastic-flow network, Computers \& Operations Research 30 (4) (2003) 567–575.
		
		\bibitem{forghani2022quickest}
		M. Forghani-elahabad, W.-C. Yeh, An improved algorithm for reliability evaluation of flow
		networks, Reliability Engineering \& System Safety (2022) 108371.
		
		\bibitem{yeh2015fast}
		W.-C. Yeh, A fast algorithm for quickest path reliability evaluations in multi-state flow net-
		works, IEEE Transactions on Reliability 64 (4) (2015) 1175–1184.
		
		\bibitem{forghani20223}
		M. Forghani-Elahabad, 3 the disjoint minimal paths reliability problem, in: Operations Re-
		search, CRC Press, 2022, pp. 35–66.
		
		\bibitem{niu2021reliability}
		Y.-F. Niu, C. He, D.-Q. Fu, Reliability assessment of a multi-state distribution network under
		cost and spoilage considerations, Annals of Operations Research (2021) 1–20.
		
		\bibitem{forghani2013simple}
		M. Forghani-elahabad, N. Mahdavi-Amiri, A simple algorithm to find all minimal path vec-
		tors to demand level d in a stochastic-flow network, in: 5-th Iranian Conference on Applied
		Mathematics, September 2–4, 2013 Bu-Ali Sina University, 2013, pp. 1–4.
		
		\bibitem{chen1990quickest}
		Y. L. Chen, Y. H. Chin, The quickest path problem, Computers \& Operations Research 17 (2)
		(1990) 153–161.
		
		\bibitem{yeh2009simpleQPR}
		W.-C. Yeh, W.-W. Chang, C.-W. Chiu, A simple method for the multi-state quickest path
		flow network reliability problem, in: 2009 8th International Conference on Reliability, Main-
		tainability and Safety, IEEE, 2009, pp. 108–110.
		
		\bibitem{el2016efficient}
		M. El Khadiri, W.-C. Yeh, An efficient alternative to the exact evaluation of the quickest path
		flow network reliability problem, Computers \& Operations Research 76 (2016) 22–32.
		
		\bibitem{forghani2020ijor}
		M. Forghani-Elahabad, N. Mahdavi-Amiri, N. Kagan, On multi-state two separate minimal
		paths reliability problem with time and budget constraints, International Journal of Opera-
		tional Research 37 (4) (2020) 479–490.
		
		\bibitem{forghani2019ress}
		M. Forghani-elahabad, N. Kagan, N. Mahdavi-Amiri, An mp-based approximation algorithm
		on reliability evaluation of multistate flow networks, Reliability Engineering \& System Safety
		191 (2019) 106566.
		
		\bibitem{forghani2016amm}
		M. Forghani-elahabad, N. Mahdavi-Amiri, An improved algorithm for finding all upper bound-
		ary points in a stochastic-flow network, Applied Mathematical Modelling 40 (4) (2016) 3221–
		3229.
		
		\bibitem{fathabadi2009determining}
		H. S. Fathabadi, M. Soltanifar, A. Ebrahimnejad, S. Nasseri, Determining all minimal paths
		of a network, Australian Journal of Basic and Applied Sciences 3 (4) (2009) 3771–3777.
		
		\bibitem{forghani2019ijrqse}
		M. Forghani-elahabad, N. Kagan, Assessing reliability of multistate flow networks under cost
		constraint in terms of minimal cuts, International Journal of Reliability, Quality and Safety
		Engineering 26 (05) (2019) 1950025.
		
	\end{thebibliography}
	%\end{document}

\end{document}